\documentclass[fleqn,twoside]{article}
\usepackage{espcrc2}

\newcommand{\AmS}{{\protect\the\textfont2
  A\kern-.1667em\lower.5ex\hbox{M}\kern-.125emS}}

\newcommand{\be}{\begin{equation}}
\newcommand{\ee}{\end{equation}}
\newcommand{\bea}{\begin{eqnarray}}
\newcommand{\eea}{\end{eqnarray}}
\newcommand{\no}{\noindent}
\newcommand{\nn}{\nonumber}

\newcommand{\Tr}{{\rm Tr\,}}
\newcommand{\Det}{{\rm Det\,}}

\newcommand{\ra}{\right\rangle}
\newcommand{\la}{\left\langle}

\newcommand{\parm}{\par\medskip}

\title{Large Mass and Chemical Potential Model: A Laboratory for QCD?}

\author{Ralf Hofmann\address[MCSD]{Institut f\"ur Theoretische Physik, Univ. Heidelberg, Heidelberg, Germany}%
              and
       Ion-Olimpiu Stamatescu$^{a,}$\address[MCSD]{Forschungsst\"atte der Evang. Studiengemeinschaft, Heidelberg}}
       
\begin{document}

\begin{abstract}
We use a model based on the hopping parameter expansion to study QCD at large
$\mu$. We find interesting behavior in the region expected to show flavor-color locking.
%\vspace{1pc}
\end{abstract}

\maketitle

\begin{figure}[h]
\vspace{6cm} 
%\center{
\includegraphics{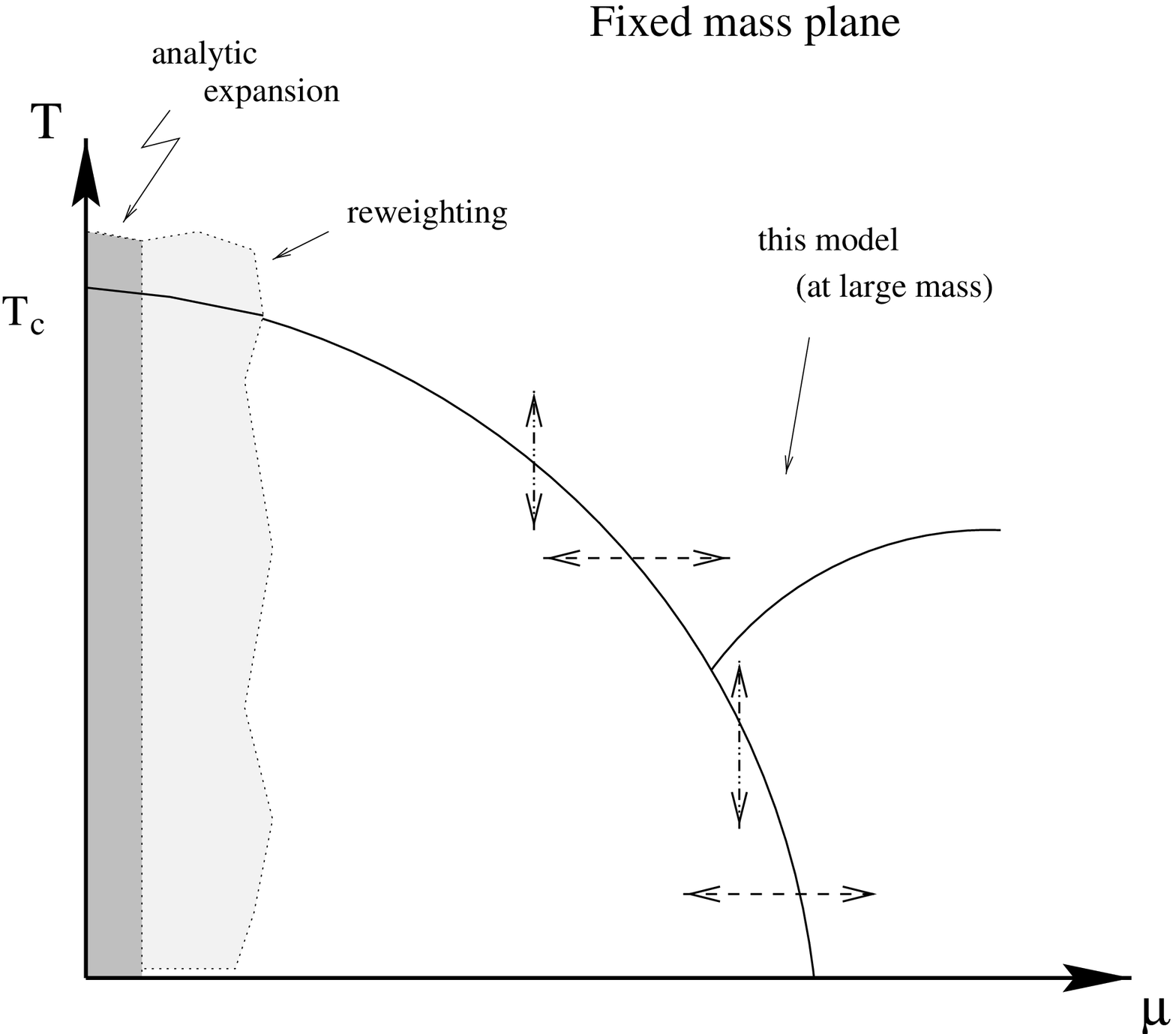} 
\includegraphics{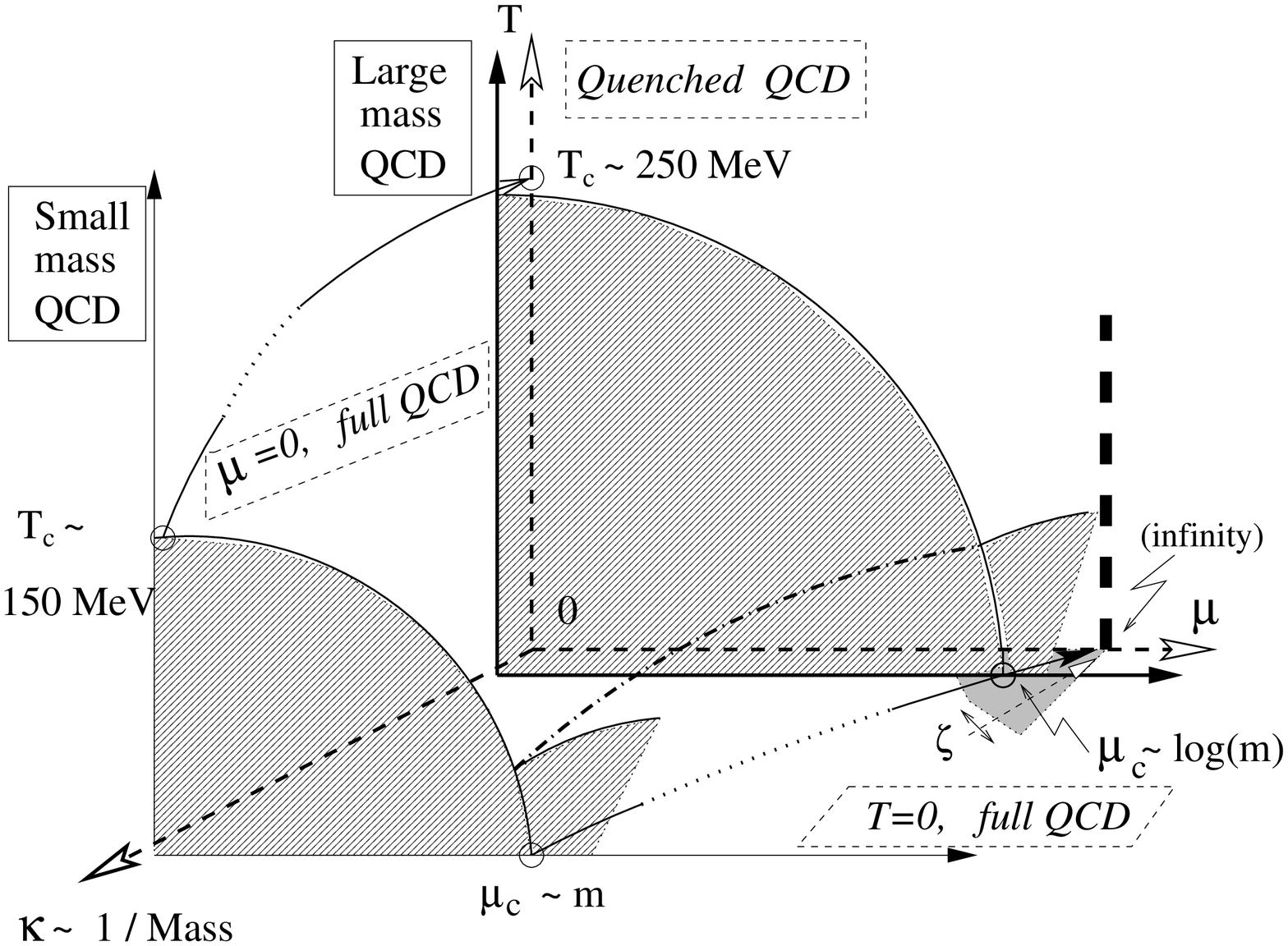} 
%}
\caption{Tentative phase diagram of QCD.} 
\label{f.kappa} 
\vspace{-0.5cm} 
\end{figure}

\no {\bf Motivation and problem setting}\ 
While the analysis for QCD at small $\mu$ and large $T$ 
gradually improves, there are no studies yet at large $\mu$ and small $T$.
Results hereto have only been obtained  in models which are not directly derivable from QCD, such as SU(2) and isovector chemical potential, or NJL.
Here we want to approach this very interesting region \cite{AlfWilczekRaj}
in a model which 
can be obtained as large mass, large chemical potential approximation
 of full QCD \cite{hdm01} and which in the static, dense quark limit 
\cite{bend}  has been proposed as quenched formulation
of QCD at finite baryon density \cite{fktre}. See Fig. \ref{f.kappa}.
For this we expand the fermion determinant in the inverse mass
to second order, which ensures a limited mobility of the quarks. 
We use Wilson fermionic action and  have for the hopping parameter:
\bea 
\kappa = 
[2(M+3+\cosh \mu)]^{-1} = [2(M_0+4)]^{-1}  \nn
\eea
\no ($M:$ bare mass, $M_0:$  bare mass at $\mu=0$). Then:
\bea
{\cal Z}_F^{[2]}({ {\kappa}}, \mu, \left\{U\right\}) 
 =   {\rm exp}\left\{-2\,  \sum_{\left\{{\vec x}\right\}}\,
\sum_{s=1}^\infty \,{{{ (\epsilon\, C
)}^s}\over s} \right. {\times} \nn 
\eea
\vspace{-0.4cm}
\bea
\left. \Tr \!\!
  \left[({\cal P}_{\vec x})^s  + \kappa^2\sum_{r,q,i,t,t'}
(\epsilon\, C)^{s(r-1)}({\cal P}_{{\vec x},i,t,t'}^{r,q})^s \right]\right\}
\!\! \label{e.corr2}
\eea
\vspace{-0.4cm}
\bea
= {\cal Z}_F^{[0]}( C,  \left\{U\right\})   
 \!\!\!\!\!\!  \prod_{{\vec x}, r,q,i,t,t'} \!\!\!\!
  \Det \left(1-(\epsilon\,C)^{r}\,\kappa^2\,
  {\cal P}_{{\vec x},{i},t,t'}^{r,q}\right)^2 , 
\nn
\eea
\vspace{-0.4cm}
\bea
C=(2\zeta)^{N_{\tau}},\ \  \zeta= \kappa\,e^{\mu} \label{e.zeta}
\eea
In the temporal gauge ($U_{n,4}=1$, except for $U_{({\vec x}, n_4=N_\tau),4} \equiv V_{\vec x}$: free) we have ${\cal P}_{\vec x} = V_{\vec x}$ and
\be
{\cal P}_{{\vec x},i,t,t'}^{r,q} = (V_{\vec x})^{r-q} 
U_{({\vec x},t),i} (V_{{\vec x}+{\hat {\i}}})^q 
U_{({\vec x},t'),i}^*
\ee
\no with
$r>q \geq 0$, $i =\pm 1, \pm 2, \pm 3$, 
$1 \leq t \leq t' \leq N_{\tau}$ ($t < t'$ for $q=0$). Here $\epsilon = 1 \, (-1)$ for p.b.c. (a.p.b.c).
For detail see \cite{hdm01}.

The hope is that the model retains some features of QCD in the 
large $\mu$ region, and that due to the simplicity of the determinant
 large statistics
reweighting methods could converge, giving  insight in the 
phase structure of QCD.

\begin{figure}[h]
\vspace{8cm} 
%\center{
\includegraphics{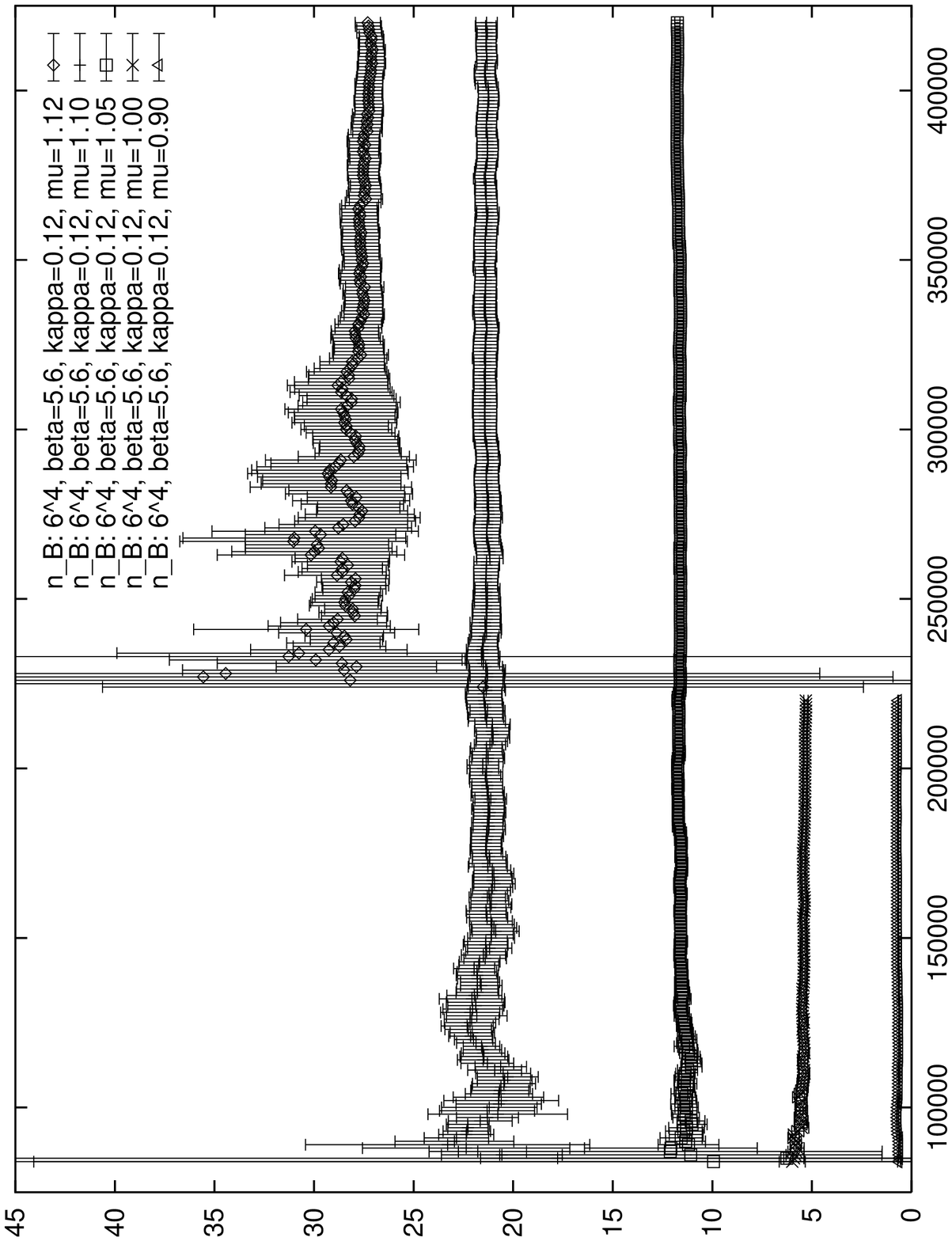} 
\includegraphics{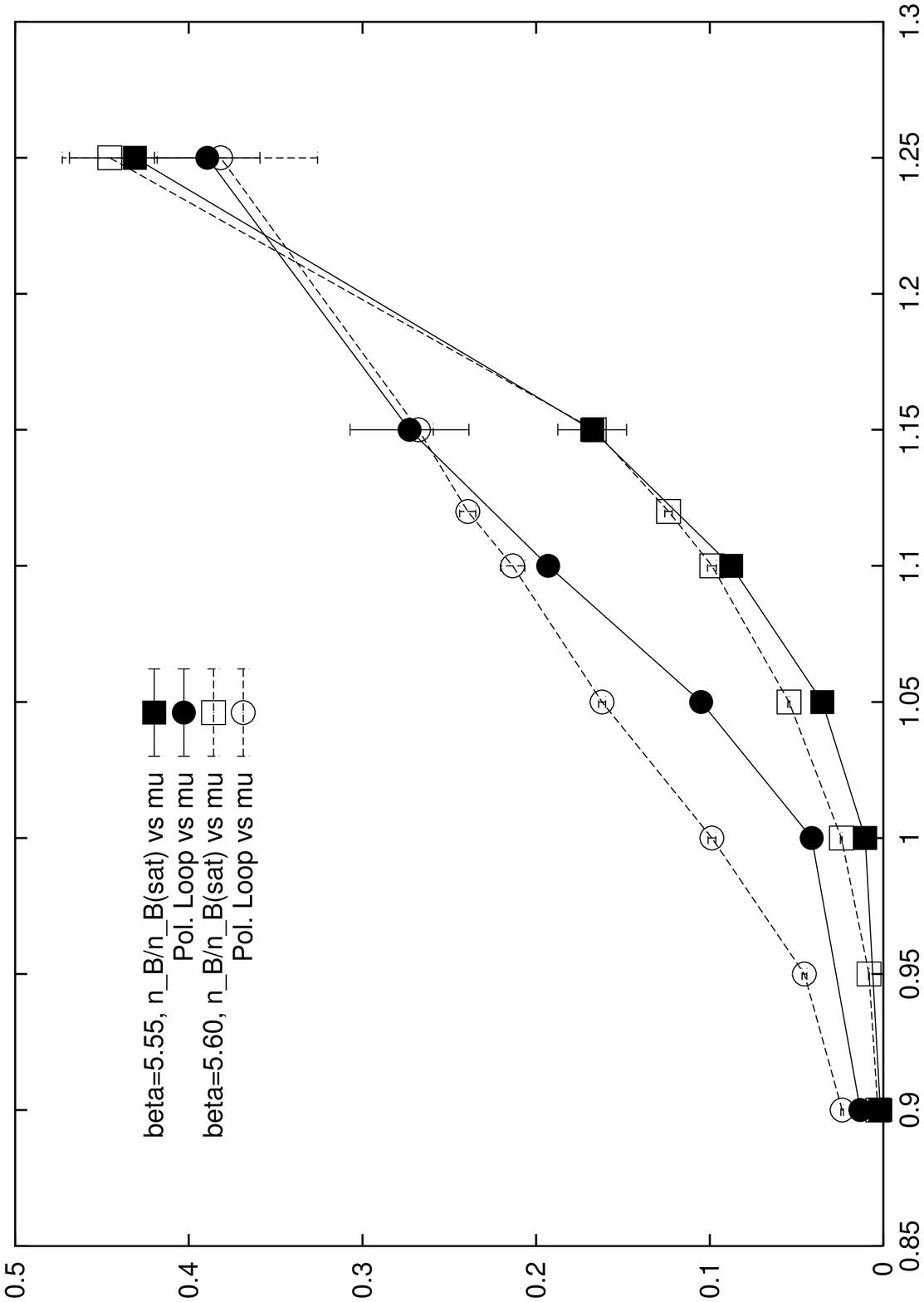} 
%}
\caption{One flavor QCD. Upper plot: iteration history for $n_B$.
Lower plot: $n_B/n_B(sat)$ and Polyakov loop vs $\mu$.} 
\label{f.cdp1} 
\vspace{-0.5cm} 
\end{figure}

\no {\bf One flavor QCD}\
The simulations used here and below are done on a $6^4$ periodic lattice
(a.p.b.c. for fermions in $\tau$-direction, Wilson plaquette action for the gauge field). 
The  algorithm uses a local 
Boltzmann factor obtained by taking out from ${\cal Z}_F^{[2]}$ the factor
\bea
\prod_{\vec x}
{\rm exp}\left\{ 2C
{\cal R}e\Tr 
  \left[{\cal P}_{\vec x}  + \kappa^2\sum_{r,q,i,t,t'}
{\cal P}_{{\vec x},i,t,t'}^{r,q} \right]\right\} \nn
\eea
compensating for this in the global reweighting (both procedures are vectorized).
We use temporal gauge fixing for easy book-keeping of the
contributions to  ${\cal Z}_F^{[2]}$. 
We first analyze one flavor QCD, some results are shown in Fig. \ref{f.cdp1}.
The convergence appears rather good, up to very large $n_B$ (baryon density over $T^3$; $n_B(sat)=216$). Both $n_B$ and Polyakov loop increase above 
$\mu \sim 1$.

\begin{figure}[h]
\vspace{9cm} 
%\center{
\includegraphics{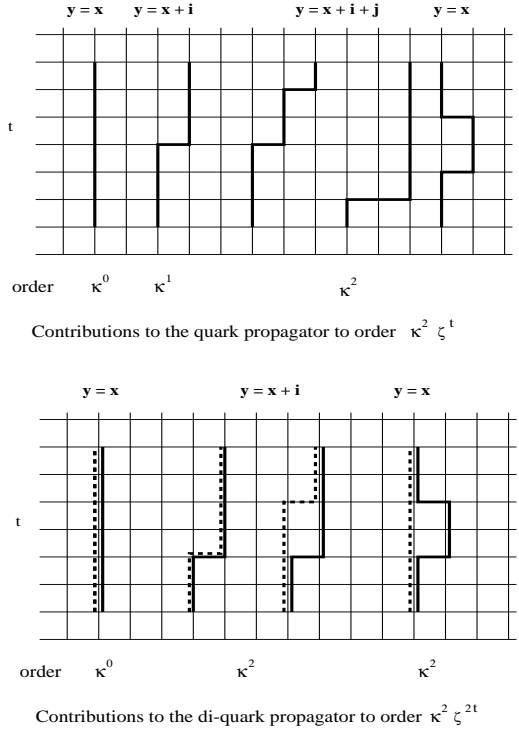} 
%}
\caption{Paths contributing to ${\cal O}(\kappa^2)$.} 
\label{f.paths} 
\vspace{-0.5cm} 
\end{figure}

\no{\bf Application to QCD with three degenerate flavors: Color-flavor locking}\
We now investigate whether a transition to a phase, where the 
ground state is characterized by a color-flavor locking condensate of 
quark bilinears \cite{AlfWilczekRaj}, takes place. Disregarding 
the possibility of parity violation in the ground state, 
such a condensate most generally is parameterized as
%***********
\be
\label{CFLC}
\la \psi^a_i C\gamma^5 \psi^b_j\ra=K_1\,\delta^a_i\delta^b_j+K_2\,\delta^a_j\delta^b_i\,
\ee
%***********
where $C=i\gamma_2\gamma_0$ is the charge conjugation matrix. 
The contraction of Dirac indices is implicit while $a\cdots h$ and $i\cdots z$ denote color and flavor 
indices, respectively. The condensate is invariant under locked global color and 
vectorial flavor rotations. To project out the 
invariants $K_1$ and $K_2$ we apply 
$\lambda\left(\delta_i^a\delta^b_j+\rho\,\delta^a_j\delta^b_i\right)$
to the condensate (\ref{CFLC}) with $(\lambda,\rho)=(-\frac{1}{36},-3)$ and 
$(\lambda,\rho)=(\frac{1}{8},-\frac{1}{3})$, respectively. 
We define the two-point correlator of color-flavor locking quark bilinears as
$\la \psi^a_i(x) C\gamma^5 \psi^b_j(x) \left[\psi^c_k(0) 
C\gamma^5 \psi^d_l(0)\right]^\dagger\ra$.
Two-point correlations of a certain combination of the invariants $K_{1,2}$ 
are obtained by applying 
%********
\be
\label{proj-cor}
\lambda(x)\lambda(0)\left(\delta_i^a\delta^b_j+\rho(x)\,\delta^a_j\delta^b_i\right)
\left(\delta_k^c\delta^d_l+\rho(0)\,\delta^c_l\delta^d_k\right) 
\ee
%********
to the above correlator after  accordingly adjusting 
$(\lambda(x),\rho(x))$ and $(\lambda(0),\rho(0))$. 
In the Euclidean formulation the result can be decomposed as 
%*********
\bea
\lambda(x)\lambda(0)\!\!\sum_{I,J=1,2}\!\!\la\psi^a_i(x)O^{I,ab}_{ij}
\psi^b_j(x)\bar{\psi}^d_l(0)\bar{O}^{J,cd}_{kl}
\bar{\psi}^c_k(0)\ra\ \nn 
\eea
\vspace{-0,5cm}
\bea
&=& O^I_{AB}\bar{O}^J_{CD} \, \left(W^{-1}_{A,C}(x,0)W^{-1}_{B,D}(x,0) \right. \nn \\
&-&\left.
W^{-1}_{A,D}(x,0)W^{-1}_{B,C}(x,0)\right)\, \det W
\label{e.corr}
\eea
%*******
(see, e.g., see for example \cite{Nikolai}), where
%*********
\bea
\label{operators}
O^{1,ab}_{ij}&=&\delta_i^a\delta^b_j\gamma_2\gamma_4\gamma_5, \ 
O^{2,ab}_{ij}=\rho(x)\delta_j^a\delta^b_i\gamma_2\gamma_4\gamma_5\,,\nonumber\\ 
\bar{O}^{1,cd}_{kl}&=&\delta_k^c\delta_l^d\gamma_5\gamma_4\gamma_2, \ 
\bar{O}^{2,cd}_{kl}=\rho(0)\delta_l^c\delta_k^d\gamma_5\gamma_4\gamma_2 \nn
\eea
Here a 'super' index $A=(\alpha,a,i)$ was 
introduced with $\alpha$ a Dirac index. Like  $\det W$ $W^{-1}$  can also
be expanded in powers of $\kappa$ \cite{hdm01}, the paths contributing to ${\cal O}(\kappa^2)$
are shown in Fig. \ref{f.paths}.

\begin{figure}[h]
\vspace{9.5cm} 
%\center{
\includegraphics{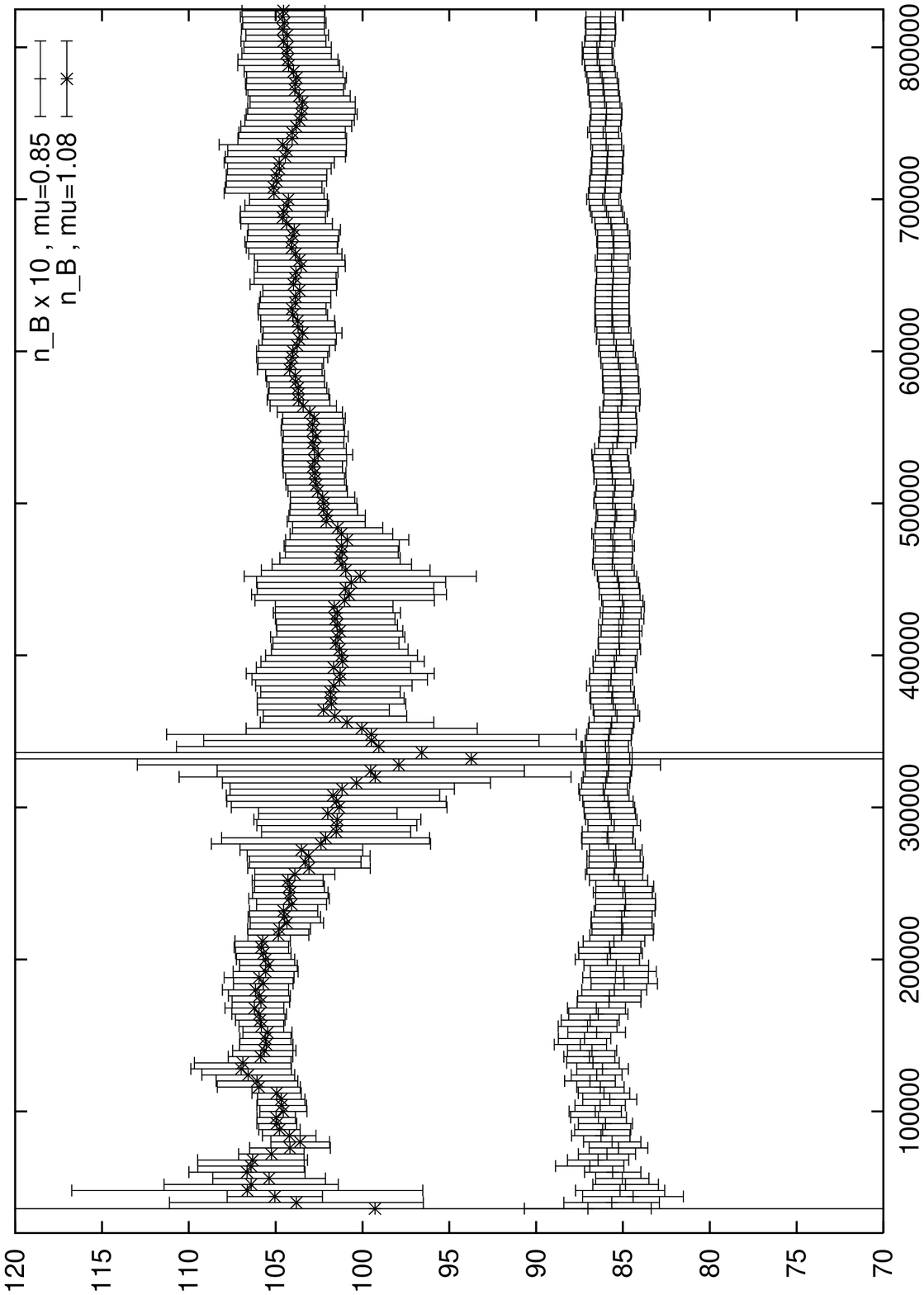} 
\includegraphics{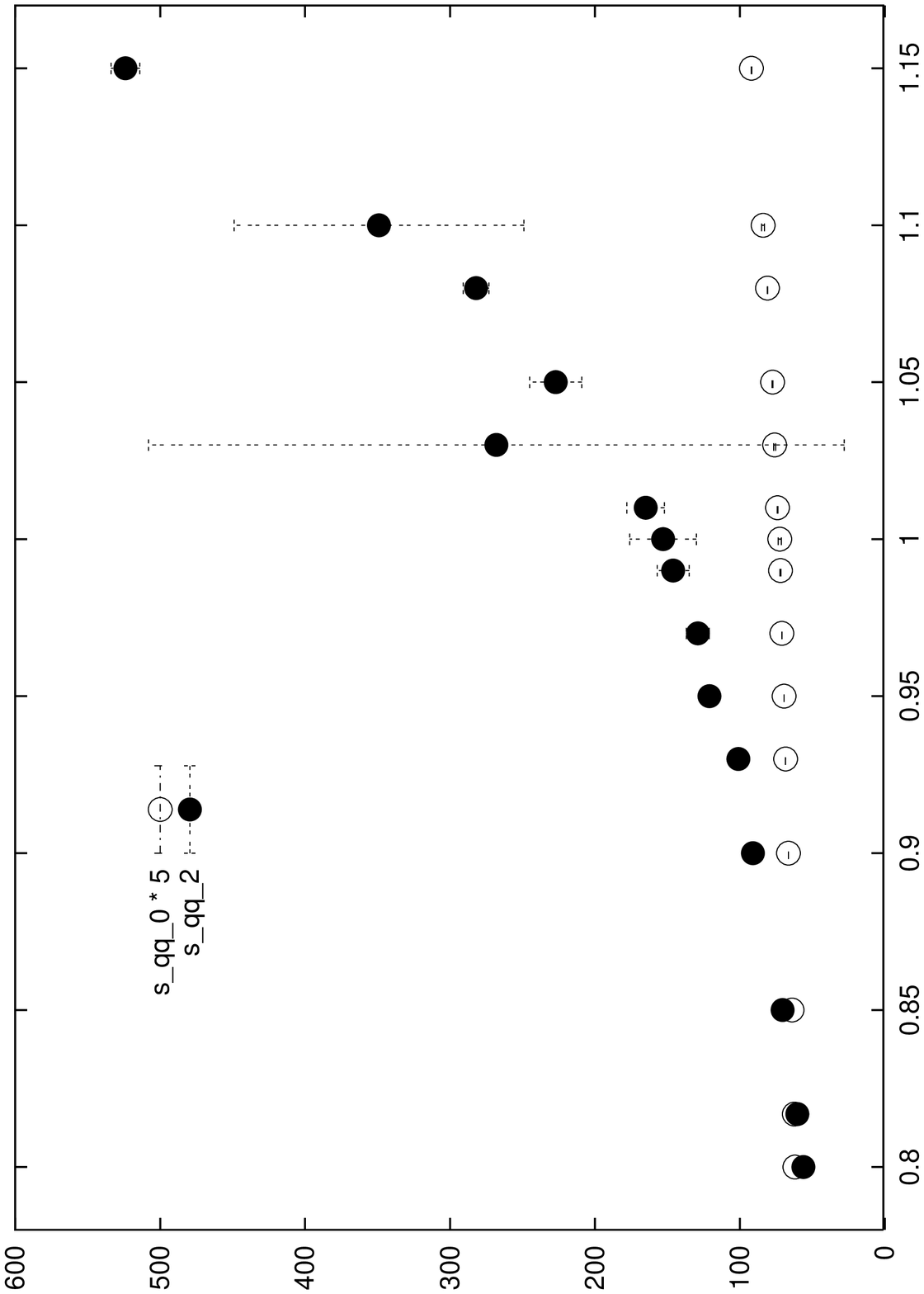} 
%}
\caption{3-flavor QCD. Upper plot:
Iteration history for $n_B$ at $\mu=0.85$ (multiplied by 10) and $1.08$.
Lower plot: ``QQ"-susceptibility to ${\cal O}(\kappa^0)$ (multiplied by 5)
and to  ${\cal O}(\kappa^2)$, vs $\mu$.} 
\label{f.cds3} 
\vspace{-0.5cm} 
\end{figure}

We have performed simulations for QCD with 3 degenerate flavors
at $\beta=5.6$ on a $6^4$ lattice
at $\kappa=0.12$ and various $\mu$ (same for all flavors). Results are shown in
Fig. \ref{f.cds3}. The general convergence is poorer 
than for the 1-flavor case at same $\mu$. In particular, the region 
around $\mu=1$ acknowledges strong fluctuations, while at larger $\mu$
the situation improves again. Here we also measure
 the ``QQ"-susceptibility obtained by integrating 
the correlator of $K_1+\xi\,K_2$ (\ref{e.corr}), where for definiteness we chose $\xi=0.5$. For this quantity we use maximal gauge fixing.
The $\kappa^2$ contribution appears essential for the
non-trivial large $\mu$ behavior (the "mobility" defined as the ${\cal O}(\kappa^2)$
fraction of $n_B$ over total $n_B$ is typically $\sim 0.3$).\parm

\no{\bf Conclusions and outlook}\ 
We have performed simulations at small $T$, large $\mu$ for a model which can be obtained 
from QCD by small order hopping parameter expansion at large $\mu$. The quarks, although
very heavy, have some limited amount of mobility. The results show  strong 
increase of the baryon density and other observables above $\mu \sim 1$ 
which may be a signal for entering a new, high density phase at small temperature.
Particularly interesting is the behavior of the di-quark correlator in 3-flavor
QCD which becomes
increasingly flat at large $\mu$, leading to a strongly increasing susceptibility. 
This may be a signal for the development of a condensate with  color-flavor locking.
Simulations on larger lattices would be essential to check this conjecture, it remains to be seen whether good convergence can be achieved in that case.\parm

\no{\bf Acknowledgments}\ The calculations have been performed on the VPP5000 of the 
Forschungszentrum Karlsruhe and the University Karlsruhe.

\bibliographystyle{prsty}

\end{document}